\documentclass[pdflatex,sn-mathphys-num,twocolumn]{sn-jnl}

\usepackage{stackengine}
\usepackage{graphicx}%
\usepackage{subfig}%
\usepackage{multirow}%
\usepackage{amsmath,amssymb,amsfonts}%
\usepackage{amsthm}%
\usepackage{mathrsfs}%
\usepackage[title]{appendix}%
\usepackage{xcolor}%
\usepackage{textcomp}%
\usepackage{manyfoot}%
\usepackage{booktabs}%
\usepackage{algorithm}%
\usepackage{algorithmicx}%
\usepackage{algpseudocode}%
\usepackage{listings}%
\usepackage{geometry}
\usepackage{lineno}
\usepackage{tabularx}
\usepackage{multicol}
\usepackage{array}
\newcolumntype{R}{>{\raggedleft\arraybackslash}X}

\theoremstyle{thmstyleone}%
\theoremstyle{thmstyletwo}%

\theoremstyle{thmstylethree}%

\begin{document}

\title[Pion emission source shape in UrQMD Au~+~Au collisions at STAR energies]{Pion emission source shape in UrQMD Au~+~Au collisions at STAR energies}



\author*[1]{\fnm{Mátyás} \sur{Molnár}}\email{molnarmatyas@student.elte.hu}

\author[1]{\fnm{Dániel} \sur{Kincses}}

\author[1]{\fnm{Yan} \sur{Huang}}

\author[1]{\fnm{Máté} \sur{Csanád}}

\affil*[1]{\orgdiv{Department of Atomic Physics}, \orgname{ELTE Eötvös Loránd University}, \orgaddress{\street{Pázmány Péter sétány 1/A}, \city{Budapest}, \postcode{1117}, \state{Hungary}, \country{Hungary}}}




\abstract{Femtoscopic measurements of two-pion Bose--Einstein correlations have established that particle-emitting sources in heavy-ion collisions are well described by Lévy $\alpha$-stable distributions, motivating systematic studies across a wide range of collision energies. In this work, we present a three-dimensional femtoscopic analysis of pion pairs in Au+Au collisions simulated with the UrQMD model for collision energies $\sqrt{s_{NN}}=3$--$27\,\mathrm{GeV}$, taking the RHIC BES-II range of collider and fixed target experiment energies for reference. 
Using Lévy-type source parameterisations, we extract the Lévy index $\alpha$, the three-dimensional radii $R_\mathrm{out}$, $R_\mathrm{side}$, $R_\mathrm{long}$, and the pair multiplicity parameter $\lambda^{*}$ (related to the correlation strength $\lambda$). We investigate their dependence on the transverse mass ($m_T$) and collision energy, along with derived quantities such as the radius difference $R_{\mathrm{diff}}^2=R_{\mathrm{out}}^2-R_\mathrm{side}^2$ and the ratio $R_\mathrm{out}/R_\mathrm{side}$. We find that $R_\mathrm{out,side,long}$ all decrease with increasing $m_T$ and increase with collision energy, consistent with collective expansion, $R_\mathrm{long}$ showing the strongest and $R_\mathrm{side}$ the weakest energy dependence. The Lévy index $\alpha$ decreases with collision energy, with a larger $m_T$-dependence towards higher energies. The $\lambda^{*}$ parameter is consistent with a constant close to unity in the absence of pions from long-lived resonances. These results provide a baseline for future comparisons with experimental measurements from the STAR Collaboration, contributing to constraints on the QCD phase diagram.}

\keywords{Lévy-femtoscopy; STAR experiment; UrQMD, QCD phase diagram, Monte Carlo event generators}




\maketitle

\section{Introduction}\label{sec1}
Femtoscopy is a sensitive tool for studying the properties of strongly interacting matter, namely the quark-gluon plasma (QGP) created in high-energy heavy-ion collisions~\citep{niida2021signatures}. This experimental technique works via the measurement of quantum-statistical momentum correlations at small relative momenta, and is commonly used in high-energy physics since the discovery of the GGLP effect~\citep{goldhaber1959pion,goldhaber1985gglp}.  This is closely related to the HBT effect discovered in astronomy~\citep{Goldhaber1990TheGE}, hence in high-energy physics it is often referred to as HBT correlation. Femtoscopy has many areas, such as two-particle and many-particle correlations~\citep{manyparticlecorr} of identical particles, and correlations of non-identical particles (also known as interaction femtoscopy~\citep{Fabbietti:2020bfg}).

One of the main aims of femtoscopy is to explore the space-time geometry of the particle-emitting source. While its shape was commonly assumed to be Gaussian~\citep{csorgo1994quantum,akkelin1995hbt} across collision energies and particle species, more recent measurements indicate that using Lévy-stable source distributions results in a better description of the measured correlation functions over a broad range of SPS~\citep{brownSPS,Adhikary2023SPS,Porfy2024aSPS,Porfy:2025sev}, RHIC~\citep{Adare2018RHIC,Kovacs2023RHIC,Mukherjee2023RHIC, Abdulameer2024RHIC, Kincses2024RHIC}, and LHC~\citep{Tumasyan2024LHC,Korodi2023LHC} energies. This description is even more justified when looking at emission profiles exhibiting long tails~\citep{phenixlongtails}, as investigated also in several phenomenological studies ~\citep{Kincses2025,csanad2025investigating,Nagy2023,KORODI2023138295,kincses2020,Kincses:2025iaf,kincses20253DEPOS3}.

Using a three-dimensional (3D) Lévy femtoscopy method, the Lévy index of stability $\alpha$, source radii, and correlation strength can be explored, with the potential to reveal signatures of the quantum-chromodynamics (QCD) critical point~\citep{Csorgo:2005it,csanad2025investigating}. Long-range correlations near the critical point may lead to a non-monotonic energy dependence of the $\alpha$ parameter. This motivates us to analyze the energy dependence of the Lévy distribution parameters fitted to the pair source distributions simulated using the UrQMD event generator framework. The results provide a theoretical baseline for a non-critically behaving source, with a few limitations. One of the most important ones is that our results exclude an explicit hydrodynamic QGP evolution.

\section{Methods}\label{sec2}
In this work, we used the UrQMD event generator version 3.4. The UrQMD model is a microscopic transport theory that utilizes stochastic binary scatterings, color string formation, and resonance decays~\citep{bass1998microscopic,bleicher1999relativistic}. The model is intended to work in the SIS energy region ($\sqrt{s_{NN}}\approx 2$ GeV), but is also often applied in the region of the RHIC Beam Energy Scan (up to $\sqrt{s_{NN}}\approx 27$ GeV)~\cite{sahoo2024,Mitrovski:2008hb}. With a hybrid mode utilizing hydrodynamic calculations, it can even be applied at the top RHIC energy or at LHC energies~\citep{Petersen:2008dd,Petersen:2011sb}. The model is generally in-line with Au+Au collision data at RHIC energies in terms of particle yields and transverse momentum distributions~\citep{Mitrovski:2008hb,Bratkovskaya:2004kv} and is often used as a hadronic baseline~\citep{STAR:2021iop,STAR:2021fge,STAR:2021yiu,STAR:2025zdq}, in comparison with other hybrid models and experimental data across the beam energy scan range.


The event generator was used in the default hadronic cascade mode, thus providing no explicit hydrodynamic evolution or utilization of an equation of state. An interesting question is whether long-tail behavior is affected by providing no hydrodynamic phase, in contrast to other models that include one (such as our previous study with the EPOS4 model in Ref.~\citep{HUANG2026140423}).

We simulated $100\,000$ lower-energy ($\sqrt{s_{NN}}=$3.0, 3.2, 3.5, 3.9, 4.2, 4.5, 5.2, 6.2 and 7.2 ~GeV) and $10\,000$ higher-energy (7.7, 9.2, 11.5, 14.5, 19.6, 27.0~GeV) collision events in the impact parameter range of 0~fm to 4.73~fm, approximately corresponding to the 0--10\% centrality class. For lower energies, a higher number of events is required to achieve a suitable statistical precision. The impact parameter limit was chosen in accordance with a previous analysis~\citep{HUANG2026140423}, where the EPOS4 predefined internal impact parameter limits were found to be consistent with the charged particle multiplicity-based centrality definition -- the same methodology STAR uses for centrality definition, via the number of charged tracks~\citep{star2008}. We analyzed the freeze-out coordinates of pions in these events, as described below.

\subsection{Pair distance distributions}
\label{sec:extract_drho}
In contrast to the experiment, the pair distance distribution $D(\vec\rho)$ can be directly obtained in simulations using the freeze-out coordinates of identically charged pions. The three-dimensional spatial separation vector $\vec\rho$ is obtained in the Longitudinal Co-Moving System (LCMS), and calculated using the Bertsch--Pratt coordinates~\citep{Pratt1986,Bertsch1988} ``out'', ``side'', and ``long'', as discussed also in Refs. ~\citep{Kincses2025,HUANG2026140423}:
\begin{align}
    \rho_{\mathrm{out}}^{\mathrm{LCMS}} = &r_x \cos\varphi + r_y \sin\varphi - \nonumber\\
 &- \frac{K_T}{K_0^2 - K_z^2} \left( K_0 t - K_z r_z \right), \\
    \rho_{\mathrm{side}}^{\mathrm{LCMS}} = &- r_x \sin\varphi + r_y \cos\varphi, \\
    \rho_{\mathrm{long}}^{\mathrm{LCMS}} = &\frac{K_0 r_z - K_z t}{\sqrt{K_0^2 - K_z^2}}.
\end{align}
Here, $(t,r_x,r_y,r_z)$ are the particle coordinates in the laboratory frame, $K_0$ and $(K_x,K_y,K_z)$ denote the temporal and spatial components of the pair's average four-momentum $K=(p_1+p_2)/2$, while $K_T=\sqrt{K_x^2+K_y^2}$ is the pair transverse momentum, and the azimuthal angle is defined via $\cos\varphi = K_x/K_T$. In other words, the ``long'' direction is the direction of the beam, ``out'' is defined by the average pair momentum, and ``side'' is perpendicular to the previous two.

In experiments, the momentum correlation function can be measured, and it is connected to the pair source via the modified Yano-Koonin formula~\citep{Adare2018RHIC}:
\begin{equation}
    C(\vec{q}) = \int D(\vec{\rho}) \, \bigl|\psi_{\vec{q}}(\vec{\rho})\bigr|^2 \, \mathrm{d}^3 \vec{\rho}.
\end{equation}

For a Lévy-stable source, as used in this analysis, the pair distance distribution can be expressed as~\citep{Nolan2013LevyEq}
\begin{equation}\label{eq:levydistr}
D(\vec{\rho}){=}\mathcal{L}(\alpha, 2^{\frac{\alpha}{2}}{\cdot}R^2; \vec{\rho}){=}\int \frac{\mathrm{d}^3\vec{\zeta}}{(2\pi)^3} \, e^{i \vec{\zeta} 
\cdot \vec{\rho}} \, e^{-\left| \vec{\zeta}^{\,T} R^2 \vec{\zeta} \right|^{\frac{\alpha}{2}}},
\end{equation}
where the superscript $T$ denotes the transpose, $R^2 = \mathrm{diag}\left(R_{\mathrm{out}}^2, R_{\mathrm{side}}^2, R_{\mathrm{long}}^2\right)$ denotes the diagonal matrix of the Lévy scale parameters, neglecting the cross-terms~\footnote{The factor $2^{\alpha/2}$ is a convention used in order to match the experimental correlation function, where the radii describe single-particle homogeneity lengths.}. In contrast to the experimental case, where the Coulomb interactions significantly distort the correlation function at low relative momentum and need to be addressed~\cite{Nagy:2026mvg}, this treatment is not applicable, since we directly obtain the freeze-out coordinates.

\begin{figure*}[!htpb]
    \centering
    \includegraphics[width=0.95\linewidth]{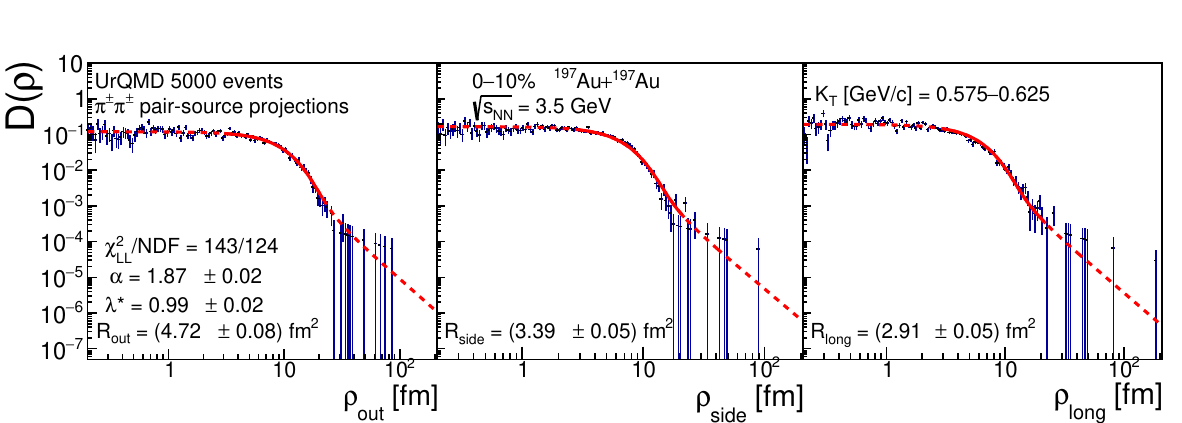}
    \caption{An example fit (red line) to  projected $D(\rho^{LCMS}_{\mathrm{out}})$, $D(\rho^{LCMS}_{\mathrm{side}})$, $D(\rho^{LCMS}_{\mathrm{long}})$ distributions (blue points) in Au+Au collisions of 0--10\% centrality at $3.5\,\mathrm{GeV}/c$, in $K_T$ range $[0.575,\,0.625]\,\mathrm{GeV}/c$. The continuous part of the red line indicates the fit range, and the dashed line is therefore an extrapolation. For this particular collision energy, 5000 events were merged for each fit. The fitted distribution corresponds to the core-core source.}
    \label{fig:examplefit}
\end{figure*}
Following the methodology of Refs.~\citep{Kincses2025,kincses20253DEPOS3,HUANG2026140423}, in this analysis, instead of using an angle-averaged distance distribution $D(\rho) = \frac{1}{4\pi} \int D(\vec{\rho}) \, \mathrm{d}\Omega$, we calculated the one-dimensional projections of the three-dimensional $D(\vec\rho)$ distribution as follows:
$$D(\rho_{\mathrm{out}}^{\mathrm{LCMS}}),\quad D(\rho_{\mathrm{side}}^{\mathrm{LCMS}}),\quad D(\rho_{\mathrm{long}}^{\mathrm{LCMS}}),$$
and fitted them simultaneously with the projections of three-dimensional Lévy distributions. In this way, the key source parameters, such as Lévy index $\alpha$, source radii or sizes $R_\mathrm{out}$, $R_\mathrm{side}$, $R_\mathrm{long}$ could be extracted, as well as the pair source strength parameter $\lambda^{*}$. A key difference to studies of Ref.~\citep{Kincses2025,kincses20253DEPOS3,HUANG2026140423} is that UrQMD does not perform weak decays (such as those of $K^0_S$, the $\eta$ meson, and $\Sigma$, $\Omega$, $\Xi$ baryons). The additional ${}^{*}$ in the notation of $\lambda$ thus refers to the fact that in UrQMD, these decays are not included, leading to ``missing'' pions in the simulation from these particles. This affects the applicability of the core-halo picture~\citep{Csorgo1996,Alt2000}, as detailed in Ref.~\citep{kincses20253DEPOS3}, Sec. 2.1. In presence of a halo, the intercept parameter $\lambda$ is defined through
\begin{equation}
    C(\vec q) = 1-\lambda + \int D_{\text{(core-core)}}(\vec\rho)\left|\Psi_{\vec q}(\vec\rho)\right|^2\mathrm{d}^3\vec\rho.
\end{equation}
However in our case the source consists almost only of the \emph{core} part, composed of primordial pions and decays of short-lived resonances ($\rho$, $\Delta$, $K^{*}$, $\omega$, $\phi$), as no longer-lived resonances contribute to the \emph{halo}. Thus we expect $\lambda^{*}$ to be close to unity. An example for such a fit is shown in Fig.~\ref{fig:examplefit}.

Similarly to Ref.~\citep{kincses20253DEPOS3}, to quantify the goodness of fit and hence the reliability of the extracted source parameters, a log-likelihood method was used, utilising \texttt{ROOT}'s~\citep{ANTCHEVA20111384} \texttt{ROOT::Math::Minimizer} class. Note that $\vec\rho$ distributions were calculated in distinct bins of transverse mass $m_T=\sqrt{m_\pi^2+K_T^2}$, where $m_\pi$ is the mass of the investigated charged pions. For detailed derivations, see Refs.~\citep{Kincses2025,kincses20253DEPOS3}. Identical charged pion pairs ($\pi^-\pi^-$ and $\pi^+\pi^+$) were selected for the analysis, with track cuts typical for STAR kinematic settings: tracks in the transverse momentum range ${0.15<p_T~[\mathrm{GeV}/c]<1.0}$ and pseudorapidity $|\eta|<1.0$ were allowed.

\subsection{Systematic uncertainty variations}
Systematic uncertainties were estimated by varying the following analysis conditions relative to the default configuration, similarly to Ref.~\citep{HUANG2026140423}.

\textbf{(1) Relative momentum selection:} In experimentally measured quantum-statistical correlations, the signal appears only at limited values of the pair momentum difference $Q_{\mathrm{LCMS}}$. To align with the experiment, an upper limit was utilized based on the $\sqrt{m_T}$ scaling of femtoscopic correlation widths~\citep{Adare2018RHIC,universe3040085}, with a default value and two (``strict'' and ``loose'') systematic variations:
$$Q_{\mathrm{LCMS}}^{\mathrm{max}} = \sqrt{A\cdot m_T},$$
where $A=0.15\,\mathrm{GeV}\text{ (default)}$, $0.05\,\mathrm{GeV}\text{ (``strict'')}$, $0.25\,\mathrm{GeV}\text{ (``loose'').}$

\textbf{(2) Fit range:} In order to study the stability of the extracted source parameters, the upper limit of the fit range was varied as a function of $\sqrt{s_{NN}}$ and $K_T$. As observed in experiments and phenomenological investigations as well, the scale parameters (and thus the width of the source distribution) often exhibit a decreasing trend with $m_T$, and an increasing trend with collision energy. Based on these observations, a linear function in $K_T$ was chosen, according to the following formula:
\begin{align}
    &\rho_{\mathrm{fit}}^{\mathrm{max}}(\sqrt{s_{NN}},K_T) = \nonumber\\&\left(\rho_{\mathrm{fit,default}}^{\mathrm{max}} - \frac{K_T-0.2\,\mathrm{GeV}/c}{0.05\,\mathrm{GeV}/c}\cdot(5.0\,\mathrm{fm})\right)\times\nonumber\\
    &\times\left(1.0\pm0.25\right)\cdot\sqrt{\frac{\sqrt{s_{NN}}}{11.5\,\mathrm{GeV}}}
\end{align}
with $\rho_{\mathrm{fit,default}}^{\mathrm{max}}=80.0\,\mathrm{fm}$, and the middle factor being $1.0$ for the default, with $\pm25\%$ for the ``loose'' and ``strict'' systematic uncertainty variations. For $K_T$, the discrete bin center values were used. This variation is determined based on parameter stability studies: the results are still stable in this region but the quantitative change still can contribute to a systematic uncertainty.

\textbf{(3) Number of merged events}: Previous studies have confirmed that non-Gaussian features are not caused by event averaging, given that Lévy-shaped sources have already appeared in event-by-event studies~\citep{KORODI2023138295,Kincses:2026tfa}; therefore, in light of smaller multiplicities at lower energies, event averaging was systematically performed. We investigated pairs from the same event only, and merged pair distributions from a set of events. The systematic uncertainty arising from the choice of number of events was determined using the values listed in Table~\ref{tab:nevtavg}. Following the methodology of Ref.~\citep{kincses20253DEPOS3}, we split the total number of available events into blocks containing $N_{\rm events}$ number of events, so, e.g., for the ${\sqrt{s_{NN}} = 4.5~\rm{GeV}}$ dataset where the maximum available number was $100\,000$, and we chose $N_{\rm events}=1\,000$ we perform 100 individual fits. We then take the average of the source parameters resulting from these fits and investigate their average transverse mass dependence at each collision energy. As a ``strict'' systematic variation, a smaller $N_{\rm events}$ value was used, and in the same way, a larger value for the ``loose'' variation (denoted by $\downarrow$ and $\uparrow$ in Table~\ref{tab:nevtavg}, respectively). The large relative leaps of $N_{\rm events}$ towards smaller $\sqrt{s_{NN}}$ are justified, as the pion pair statistics per event decrease dramatically with decreasing collision energies.

\begin{table}[h]
\normalsize
\caption{Systematic variations of $N_{\rm events}$ values, i.e. number of events averaged for each fit. We refer to the larger value as the ``loose'' and the smaller value as the ``strict'' systematic uncertainty variation.\label{tab:nevtavg}}
\centering
\begin{tabularx}{\columnwidth}{RRRR}
    \toprule
    $\sqrt{s_{NN}}$ & \multicolumn{3}{c}{$N_{\rm events}$} \\
    {[GeV]}         & Default & $\downarrow$ & $\uparrow$ \\
    \midrule
    3.0  & 20\,000 & 10\,000 & 25\,000 \\
    3.2  & 10\,000 & 5\,000  & 20\,000 \\
    3.5  & 5\,000  & 1\,000  & 10\,000 \\
    3.9  & 5\,000  & 1\,000  & 10\,000 \\
    4.2  & 1\,000  & 500     & 5\,000  \\
    4.5  & 1\,000  & 500     & 5\,000  \\
    5.2  & 500     & 200     & 1\,000  \\
    6.2  & 500     & 200     & 1\,000  \\
    7.2  & 500     & 200     & 1\,000  \\
    7.7  & 500     & 200     & 1\,000  \\
    9.2  & 200     & 100     & 500     \\
    11.5 & 200     & 100     & 500     \\
    14.5 & 200     & 100     & 500     \\
    19.6 & 100     & 50      & 200     \\
    27.0   & 100     & 50      & 200     \\
    \bottomrule
\end{tabularx}
\end{table}

In this analysis, we used asymmetric systematic uncertainties. If the systematic uncertainty variation resulted in a value above the default, it contributed to the upward systematic uncertainty, and vice versa, if it resulted in a value below the default. To summarize, in Table~\ref{tab:sysunc} we list the average relative systematic uncertainties from each source and their total contribution for selected energies. For a comprehensive list at all energies, see the Appendix, in particular Table~\ref{tab:sysunc_full}.

\begin{table*}[!htpb]
\caption{Average relative systematic uncertainties of the emission source parameters from different uncertainty sources at three selected collision energies, in percent. Note that we refer to the larger $N_{\mathrm{events}}$ value as the ``loose'' and the smaller value as the ``strict'' systematic uncertainty variation.\label{tab:sysunc}}
		\begin{tabular}{ccccccc}
\toprule
$\sqrt{s_{NN}}$ & \textbf{Source} & $\alpha$ & $\lambda^{*}$ & $R_{\mathrm{out}}$ & $R_\mathrm{side}$ & $R_\mathrm{long}$\\
\midrule
\multirow[m]{4}{*}{3.0 GeV} & $Q_\mathrm{LCMS}^\mathrm{max}$ &  +0.35 / -0.65 &  +0.25 / -0.14 &  +1.07 / -1.22 &  +2.26 / -2.38 &  +3.39 / -3.90\\
 & $\rho_\mathrm{fit}^\mathrm{max}$ &  +0.55 / -1.83 &  +0.93 / -0.29 &  +0.51 / -0.10 &  +0.23 / -0.30 &  +0.53 / -1.54\\
 & $N_{\rm events}$ &  +0.02 / -0.14 &  +0.04 / -0.01 &  +0.00 / -0.04 &  +0.01 / -0.02 &  +0.02 / -0.12\\
 & Total &  +0.75 / -1.98 &  +0.99 / -0.34 &  +1.36 / -1.26 &  +2.27 / -2.41 &  +3.43 / -4.37\\
\midrule
\multirow[m]{4}{*}{11.5 GeV} & $Q_\mathrm{LCMS}^\mathrm{max}$ &  +0.17 / -0.94 &  +0.79 / -0.06 &  +0.56 / -0.68 &  +2.19 / -2.54 &  +4.23 / -5.38\\
 & $\rho_\mathrm{fit}^\mathrm{max}$ &  +0.22 / -0.17 &  +0.06 / -0.08 &  +0.14 / -0.09 &  +0.04 / -0.03 &  +0.05 / -0.09\\
 & $N_{\rm events}$ &  +0.06 / -0.30 &  +0.15 / -0.03 &  +0.01 / -0.05 &  +0.02 / -0.05 &  +0.05 / -0.17\\
 & Total &  +0.33 / -1.07 &  +0.83 / -0.13 &  +0.60 / -0.70 &  +2.19 / -2.54 &  +4.23 / -5.39\\
\midrule
\multirow[m]{4}{*}{27.0 GeV} & $Q_\mathrm{LCMS}^\mathrm{max}$ &  +0.14 / -1.11 &  +0.49 / -0.05 &  +0.53 / -0.55 &  +2.20 / -2.66 &  +4.19 / -5.14\\
 & $\rho_\mathrm{fit}^\mathrm{max}$ &  +0.11 / -0.20 &  +0.06 / -0.03 &  +0.08 / -0.07 &  +0.02 / -0.07 &  +0.04 / -0.08\\
 & $N_{\rm events}$ &  +0.10 / -0.32 &  +0.13 / -0.04 &  +0.02 / -0.04 &  +0.04 / -0.11 &  +0.05 / -0.18\\
 & Total &  +0.30 / -1.25 &  +0.53 / -0.09 &  +0.55 / -0.56 &  +2.20 / -2.67 &  +4.20 / -5.14\\
\bottomrule
\end{tabular}
\end{table*}


\section{Results}
We present the resulting fitted femtoscopic source parameters as functions of $m_T$ and for selected $m_T$ bins as functions of $\sqrt{s_{NN}}$.

\subsection{Source shape}
\begin{figure}[h]
    \centering
    \includegraphics[width=\linewidth]{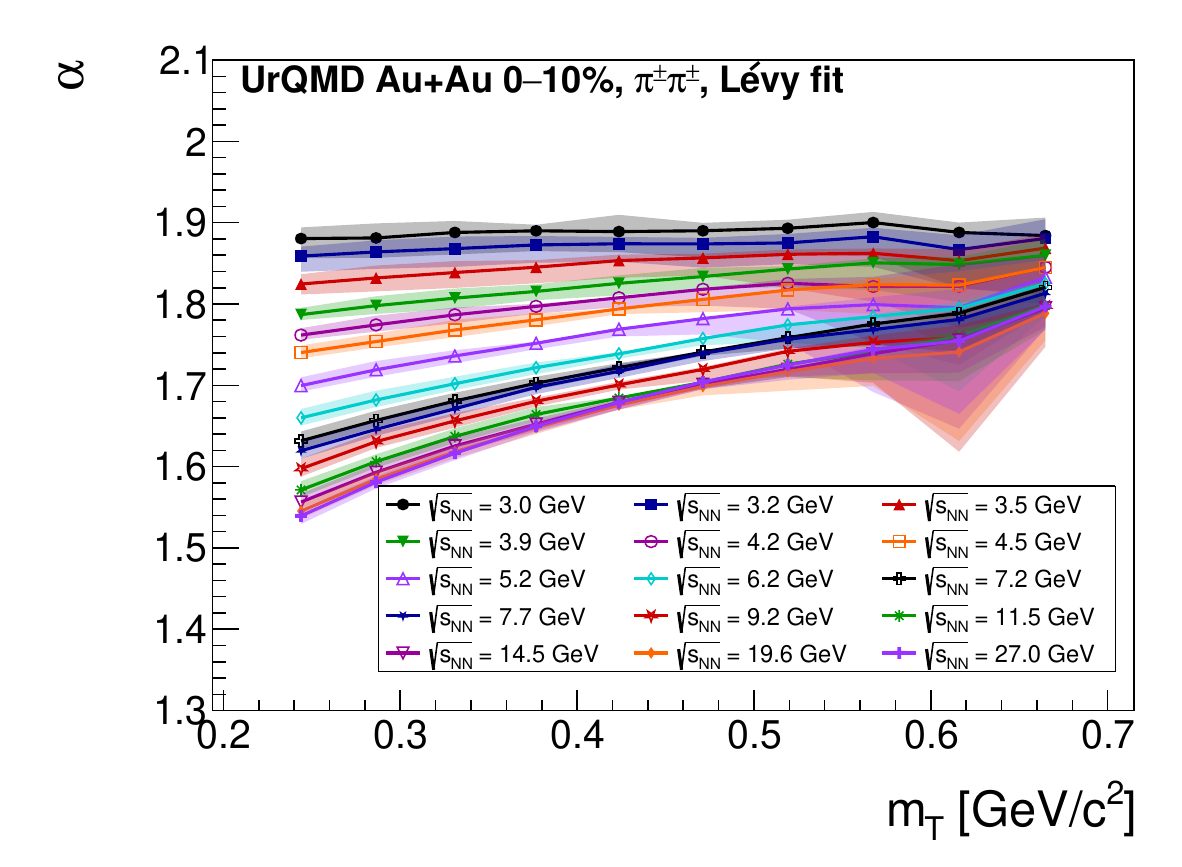}
    \caption{Lévy shape parameter $\alpha$ as a function of $m_T$. The systematic uncertainty bands include the $q_{\mathrm{LCMS}}$ cut, $\rho_{\mathrm{fit}}^{\mathrm{max}}$ and $N_{\mathrm{events-averaged}}$ uncertainty. In the last $m_T$ bins, the strict $Q_\mathrm{LCMS}$ cut variation results in probably an unrealistically large uncertainty in the downward direction, most probably the strict cut here cutting down the statistics too much and making the fit more prone to fluctuations.} 
    \label{fig:mTvsParam}
\end{figure}

Fig.~\ref{fig:mTvsParam} shows that the dependence of the Lévy index $\alpha$ on $m_T$ is accentuated significantly with increasing collision energy across all collision energies: for the lowest energies, it is consistent with a constant, and for the highest energies it shows an increasing behavior.

\begin{figure}[h]
    \centering
    \includegraphics[width=\linewidth]{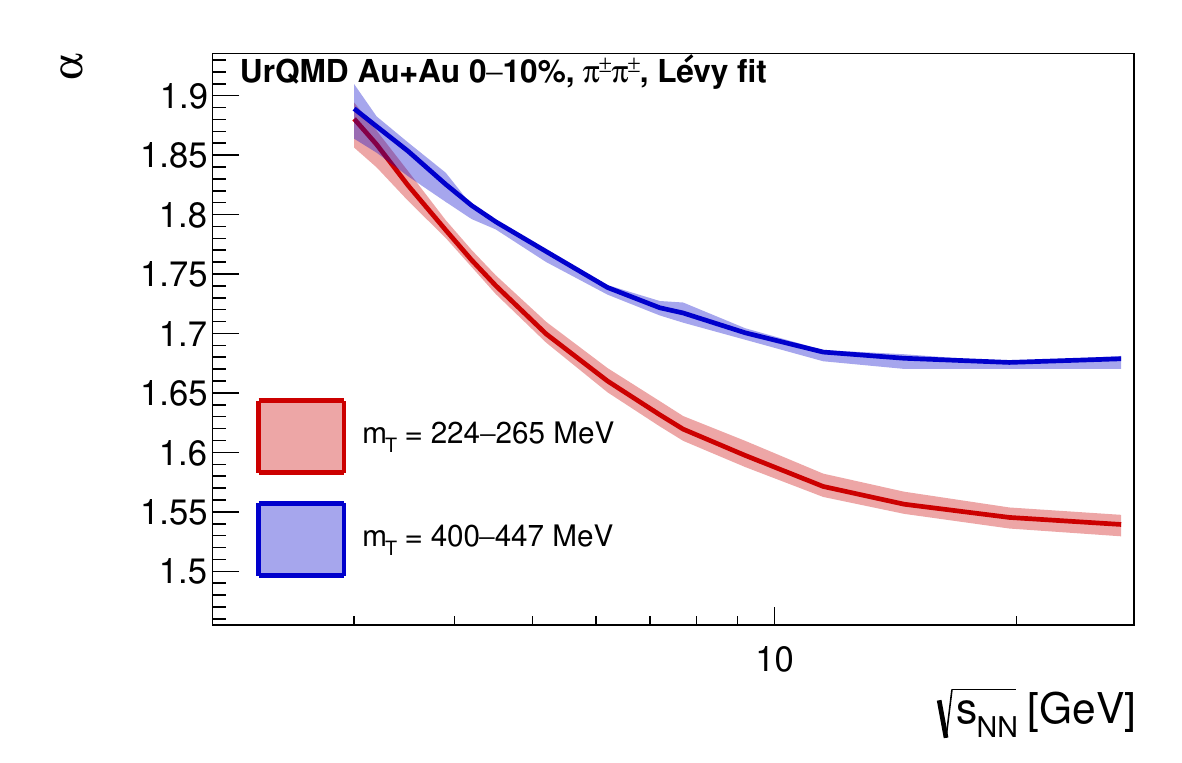}\\
    \caption{Lévy index $\alpha$ as a function of collision energy for selected $m_T$ bins corresponding to $K_T$ ranges $[0.175,0.225]$ and $[0.375,0.425]\,\mathrm{GeV}/c$}
    \label{fig:sNN_vs_alpha}
\end{figure}

In Fig.~\ref{fig:sNN_vs_alpha}, the energy dependence is presented for two $K_T$ bins: $0.175\leq K_T\,[\mathrm{GeV}/c]<0.225$ corresponding approximately to $0.224\leq m_T[\mathrm{GeV}/c^2]<0.265$ and $0.375\leq K_T\,[\mathrm{GeV}/c]<0.425$ corresponding approximately to $0.400\leq m_T[\mathrm{GeV}/c^2]<0.447$. A strong monotonic decrease was observed across the energy range, similar to the trends observed in preliminary results of the STAR experiment~\citep{Kincses2024RHIC,csanad2025investigating}.

The Lévy index $\alpha$ characterizes the shape of the emission source; $\alpha=2$ corresponds to a Gaussian, $\alpha<2$ indicates a power-law behavior, attributed to anomalous diffusion and Lévy walk in general, which includes elastic and inelastic scattering, as well as coalescence and decays. The result is thus consistent with the larger contribution of Lévy walk at lower transverse momentum and higher collision energies.

As the shape of the source is predicted to change near the QCD critical point, comparing these results with the experimentally measured $\alpha(\sqrt{s_{NN}})$ is crucial. Adding more energies to this analysis in the FAIR-CBM~\citep{Agarwal_2023} energy range may also be beneficial for future comparisons with experimental results.

\subsection{Pair source strength}
\begin{figure}[h]
    \centering
    \includegraphics[width=\linewidth]{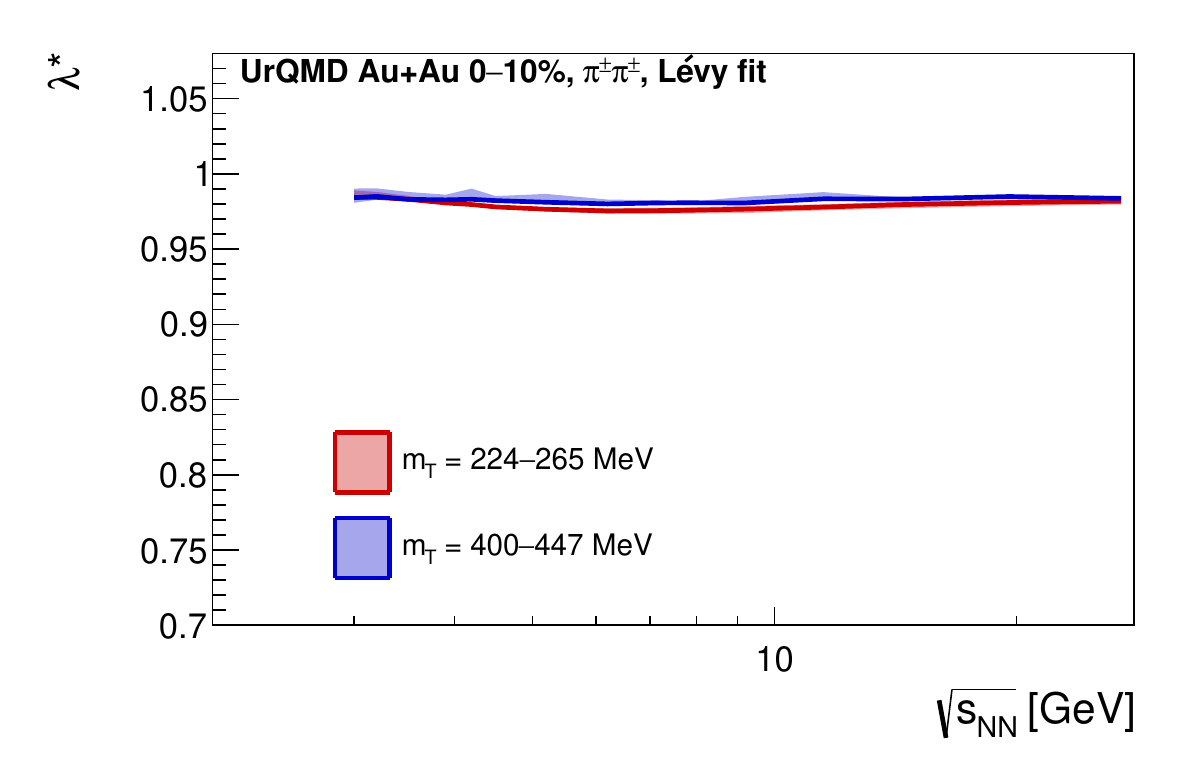}
    \caption{Correlation strength parameter $\lambda^*$ as a function of collision energy for selected $m_T$ bins corresponding to $K_T$ ranges $[0.175,0.225]$ and $[0.375,0.425]\,\mathrm{GeV}/c$}
    \label{fig:sNN_vs_lambda}
\end{figure}

The correlation strength parameter $\lambda$ is sensitive to long-lived resonances, the fraction of coherent emission, and other purity corrections~\citep{Csorgo1996}. As stated in Sec.~\ref{sec:extract_drho}, owing to the lack of simulated contributions from long-lived resonances, we used $\lambda^{*}$ in the notation for the pair source strength parameter. As shown in Fig.~\ref{fig:sNN_vs_lambda}, no clear collision energy dependence is observed apart from a mild decrease, and the value is very close to unity, as expected in this simulation. The small deviations from unity may be attributed to short-lived resonances.

\subsection{Source size}
\begin{figure*}[h]
    \centering
    \includegraphics[width=\linewidth]{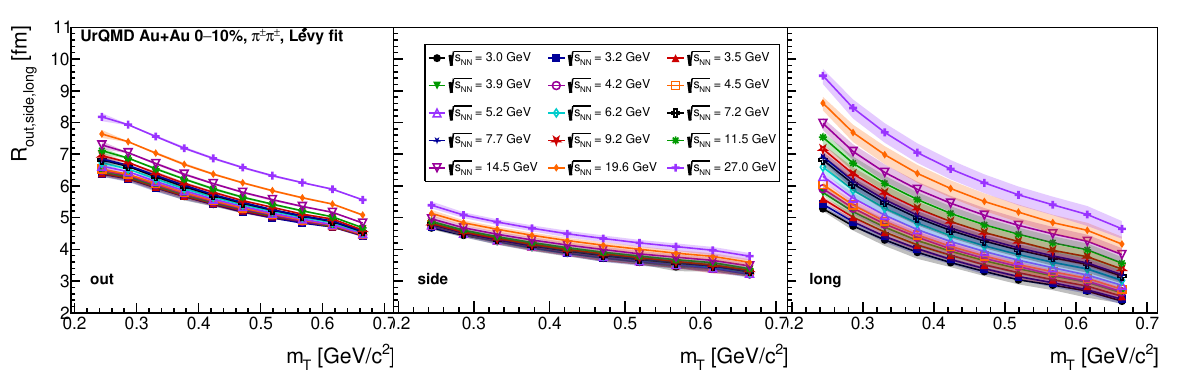}
    \caption{Lévy parameters describing the emission source size $R_{\mathrm{out}}$, $R_{\mathrm{side}}$, $R_{\mathrm{long}}$ as a function of $m_T$. The systematic uncertainty bands include the $q_{\mathrm{LCMS}}$ cut, $\rho_{\mathrm{fit}}^{\mathrm{max}}$ and $N_{\mathrm{events-averaged}}$ uncertainty.}
    \label{fig:mTvsRosl}
\end{figure*}

\begin{figure}[h]
    \stackinset{r}{15pt}{b}{15pt}{\textbf{(a)}}{%
    \includegraphics[width=\linewidth, trim={0 65pt 0 0}, clip]{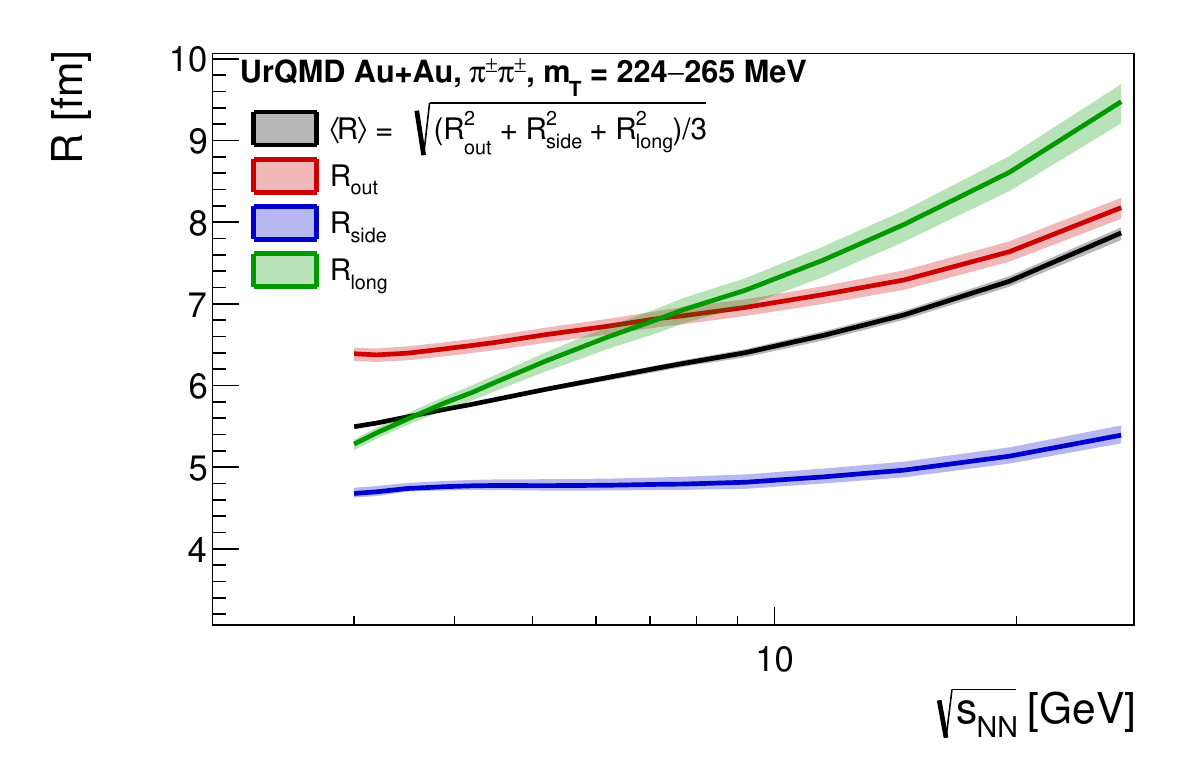}}\\
    \stackinset{r}{15pt}{b}{35pt}{\textbf{(b)}}{%
    \includegraphics[width=\linewidth, trim={0 0 0 25pt}, clip]{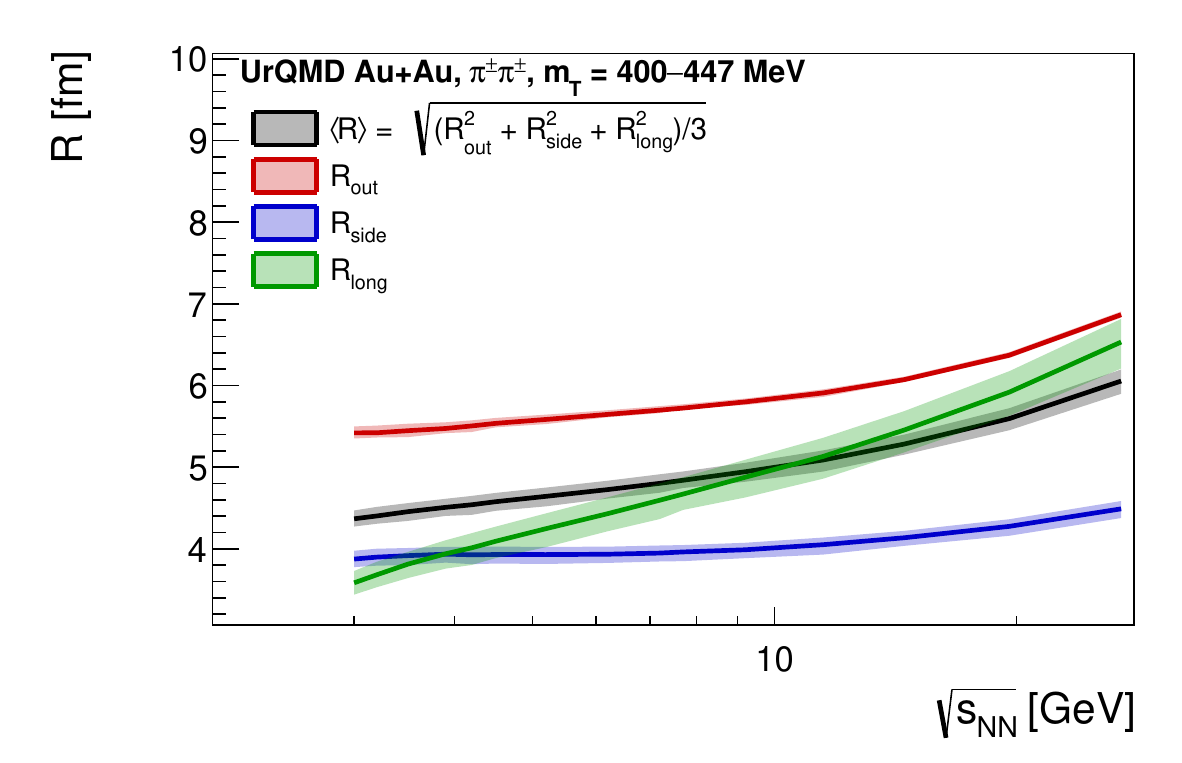}}
    \caption{Projected radii $R_{\mathrm{out}}$, $R_{\mathrm{side}}$, $R_{\mathrm{long}}$ and their average $\langle R\rangle=\sqrt{(R_{\mathrm{out}}^2 + R_{\mathrm{side}}^2 + R_{\mathrm{long}}^2)/3}$ as a function of collision energy for selected $m_T$ bins, corresponding to $K_T$ ranges (\textbf{a}) $[0.175,0.225]$ and (\textbf{b}) $[0.375,0.425]\,\mathrm{GeV}/c$.}
    \label{fig:sNN_vs_RoslR}
\end{figure}

As seen in Fig.~\ref{fig:mTvsRosl}, the extracted source radii $R_{\mathrm{out}}$, $R_{\mathrm{side}}$, and $R_{\mathrm{long}}$ show a monotonic decrease with increasing transverse momentum as expected. Furthermore, a monotonic increase with the collision energy can be observed in Fig. ~\ref{fig:sNN_vs_RoslR} for all three radii and the average $\langle R\rangle=\sqrt{(R_{\mathrm{out}}^2 + R_{\mathrm{side}}^2 + R_{\mathrm{long}}^2)/3}$. The strong increase in $R_\mathrm{long}$ is consistent with a longer system lifetime at higher $\sqrt{s_{NN}}$, and the mild increase in $R_\mathrm{side}$ and $R_\mathrm{out}$ indicates larger spatial expansion at higher energies. Overall, the $m_T$ and $\sqrt{s_{NN}}$ dependence of the source radii is in accordance with the expansion dynamics of the system, providing a stable baseline for comparison with experimental femtoscopic analyses.

\begin{figure}[h]
    \stackinset{r}{15pt}{b}{15pt}{\textbf{(a)}}{%
    \includegraphics[width=\linewidth, trim={0 65pt 0 0}, clip]{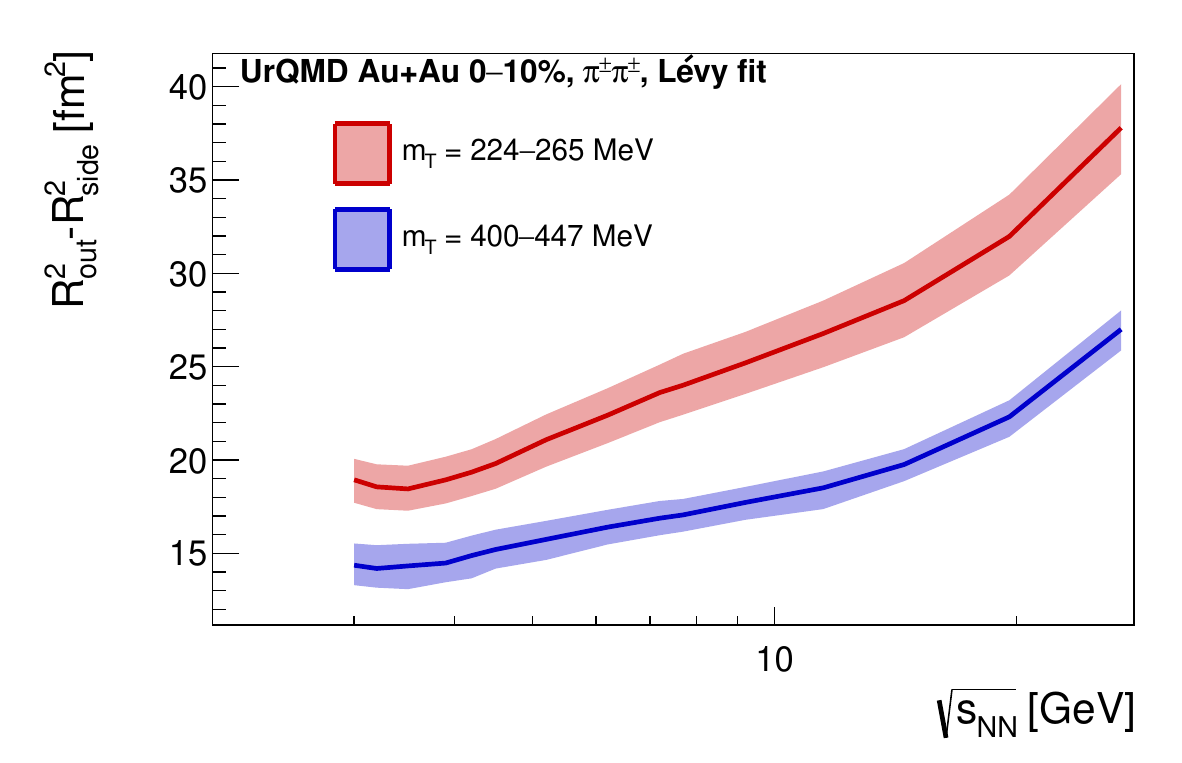}}\\
    \stackinset{r}{15pt}{b}{35pt}{\textbf{(b)}}{%
    \includegraphics[width=\linewidth, trim={0 0 0 25pt}, clip]{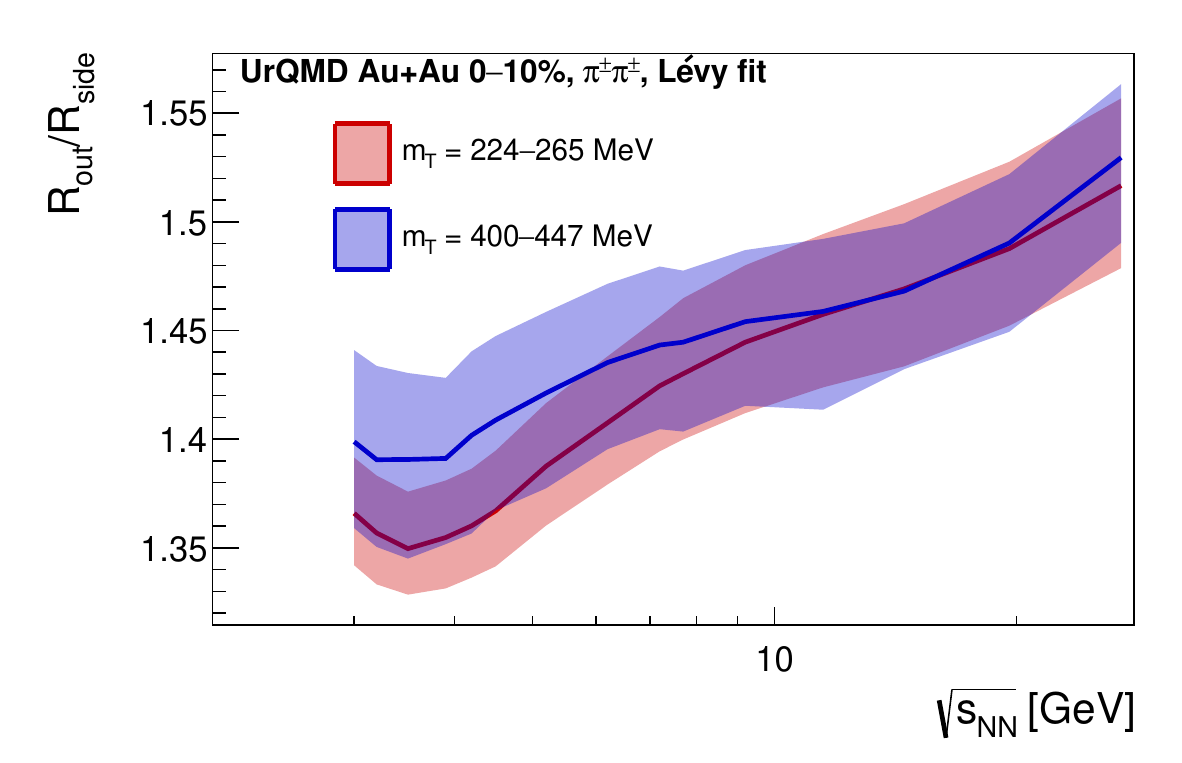}}
    \caption{Derived radius difference $R_{\mathrm{out}}^2-R_{\mathrm{side}}^2$ (\textbf{a}) and ratio $R_{\mathrm{out}}/R_{\mathrm{side}}$ and (\textbf{b}) as a function of collision energy for selected $m_T$ bins.}
    \label{fig:sNN_vs_ratio_difference}
\end{figure}

Based on Ref.~\citep{csanad2025investigating}, the radius difference ${R_{\mathrm{diff}}^2 = R_{\mathrm{out}}^{2} {-} R_{\mathrm{side}}^{2}}$ may be sensitive to potential QCD critical-point signatures. In Fig.~\ref{fig:sNN_vs_ratio_difference}(a), we show the energy dependence of this derived value as well as of the ratio $R_{\mathrm{out}}/R_{\mathrm{side}}$ in Fig.~\ref{fig:sNN_vs_ratio_difference}(b). Both quantities increase with $\sqrt{s_{NN}}$. As expected from a simple model that lacks critical behaviour, there is no clear non-monotonic trend within uncertainty. The strong deviation of the ratio from unity is most probably due to an incomplete description of the collision. The EPOS4 results shown in Ref.~\cite{HUANG2026140423} for the $R_{\mathrm{\text{out}}}/R_{\mathrm{\text{side}}}$ ratio show an opposite trend from the UrQMD results in the overlapping energy region of 7.7 to 27 GeV:  while the EPOS4 ratio decreases with increasing $\sqrt{s_{NN}}$, from about 1.52 to 1.44, the UrQMD ratio increases from about 1.43 to 1.51 -- differing by 0.08–0.09 at the edges, or approximately 5--6\% relatively. The opposite energy dependence of $R_{\rm out}/R_{\rm side}$ may reflect differences in space-time evolution and emission duration between UrQMD and EPOS4. This effect might be due to the aforementioned lack of a hydrodynamic phase with an equation of state in the UrQMD, as hydrodynamics has more impact at higher collision energies. Therefore, the results in this paper, in contrast to the EPOS4, can be considered as a hadronic baseline. Furthermore, in experiment, even in the absence of critical phenomena, a maximum may be expected in case of an onset of hydrodynamic behavior, based on the shown predictions from UrQMD and EPOS4.

\section{Discussion and Conclusions}
In this work, we presented a three-dimensional Lévy femtoscopic analysis of pion-emitting sources in UrQMD 0--10\% central Au+Au collisions at $\sqrt{s_{NN}}=3$--$27\,\mathrm{GeV}$, covering the RHIC BES-II energy range. We calculated the spatial pair distance distributions $D(\vec\rho)$ for identical charged pions in the LCMS frame. Their one-dimensional projections along the Bertsch--Pratt coordinates were simultaneously fitted using a Lévy distribution. The extracted source parameters -- Lévy index $\alpha$, intercept parameter $\lambda^{*}$, and the femtoscopic radii $R_\mathrm{out}$, $R_\mathrm{side}$, $R_\mathrm{long}$ -- establish a hadronic transport baseline free of critical-point effects, against which future experimental data can be tested.

The phenomenologically most significant finding of this work is the strong, monotonic decrease of the Lévy index $\alpha$ with increasing collision energy. At the lowest simulated energies the $m_T$ dependence of $\alpha$ is consistent with a constant, whereas it becomes markedly accentuated at higher $\sqrt{s_{NN}}$, suggesting that at higher energies and lower $m_T$ the contribution of resonances (contributing to the source function below $\sim100~\rm{fm}$) would be more pronounced. The values of $\alpha$ remain well below 2 across all energies and $m_T$ bins, confirming that a Gaussian source description is inadequate in this kinematic regime. Crucially, since UrQMD does not incorporate a QCD phase transition or critical-point dynamics, the smooth, monotonically decreasing $\alpha(\sqrt{s_{NN}})$ obtained here represents the expected hadronic-transport trend in the absence of criticality. A non-monotonic energy dependence of $\alpha$ observed in STAR BES-II and other measurements would therefore be a strong indicator of dynamics beyond the hadronic baseline -- possibly related to long-range correlations near the QCD critical point~\citep{csanad2025investigating,niida2021signatures}.

It is also interesting to compare these $\alpha$ results to those obtained in Ref.~\cite{HUANG2026140423} with the EPOS4 model for Au+Au collisions in the $\sqrt{s_{NN}}=7.7-200$ GeV energy range. In the overlapping range of 7.7 to 27 GeV, an excellent quantitative agreement for $\alpha$ can be found, highlighting the strong influence of the late scattering and decay (Lévy walk) phase on the correlation function shape in these models.

The intercept parameter $\lambda^{*}$ remains close to unity across all collision energies, consistent with the expectation that, in the absence of long-lived resonance contributions, the source is dominated by the core component~\citep{Csorgo1996,Alt2000}. 

The radii $R_\mathrm{out}$, $R_\mathrm{side}$, and $R_\mathrm{long}$ all decrease monotonically with $m_T$, in qualitative agreement with the hydrodynamically motivated expectation of collective flow and the scaling of HBT radii with $m_T$~\citep{Makhlin:1987gm,Csorgo:1995bi}. The monotonic increase of all three radii with $\sqrt{s_{NN}}$ reflects the growing spatial extent and longer lifetime of the hadronic system at higher collision energies. In particular, the strong energy dependence of $R_\mathrm{long}$ is consistent with a markedly longer system lifetime at higher collision energies, while the comparatively weak energy dependence of $R_\mathrm{side}$ indicates that the transverse size of the source is less sensitive to the collision energy in this transport-only framework.

Both the ratio $R_\mathrm{out}/R_\mathrm{side}$ and the difference ${R_\mathrm{diff}^2 = R_\mathrm{out}^2 - R_\mathrm{side}^2}$ increase monotonically with $\sqrt{s_{NN}}$. This can be understood based on the  evolution of $R_\mathrm{out}$ and $R_\mathrm{side}$ as a function of collision energy, as shown in Fig.~\ref{fig:sNN_vs_RoslR}. The rather strong rise of $R_\mathrm{out}^2-R_\mathrm{side}^2$, which is larger than that typically observed experimentally, is most likely a consequence of the longer emission duration in a cascade model, due to the increased duration of hadronic scatterings in higher multiplicity systems. The ratio $R_\mathrm{out}/R_\mathrm{side}$ increases less strongly than the difference, as the ratio is not affected by the overall increase in system size with collision energy~\cite{csanad2025investigating}.

Comparing these results of radii to those obtained with EPOS4 in Ref.~\cite{HUANG2026140423} in the overlapping range of 7.7 to 27 GeV, $R_\mathrm{side}$ shows a strong quantitative agreement, the rising trend of $R_\mathrm{long}$ is also similar. On the other hand, $R_\mathrm{out}$, while also close quantitatively overall, exhibits no clear rise with collision energy in EPOS4, unlike in UrQMD. This can be explained by the influence of the hydrodynamic phase on source sizes, and the smaller effect of hadronic rescatterings, which are stronger tied to hadronic multiplicity. This also results in a pronounced difference in the out-side difference and ratio: both are decreasing with collision energy in EPOS4, while increasing in UrQMD. Thus in a realistic scenario, as relative strength of the hydro phase increases, a maximum in the out-side difference (or ratio) might be expected even in the absence of non-monotonic phenomena. In conjunction, as the sensitivity of Lévy scale differences to possible QCD critical-point signatures was highlighted in Ref.~\citep{csanad2025investigating}, investigations with different equations of state would also be important. Nevertheless, the smooth monotonic baseline established in the $3$--$27$ GeV collision energy range is essential for identifying any anomalous energy dependence in future experimental data.

Since UrQMD operates in hadronic cascade mode without an equation of state, the present results do not capture partonic degrees of freedom, hydrodynamic evolution, critical fluctuations associated with a QCD phase transition or medium modifications expected when a QGP phase is present~\citep{niida2021signatures}. These missing ingredients can modify the space-time evolution and long-range structure of the emitting source and therefore may also affect the extracted Lévy index $\alpha$; consequently, the present $\alpha(\sqrt{s_{NN}})$ dependence should be interpreted as a hadronic-transport baseline rather than a model-independent prediction for the non-critical QCD medium. Furthermore, extending the analysis to the lower energy range accessible to the FAIR-CBM experiment~\citep{Agarwal_2023} would enable a more comprehensive theoretical baseline spanning the baryon-dense region of the QCD phase diagram.

A possible subsequent step is a direct comparison of the Lévy source parameters extracted here with the forthcoming STAR femtoscopic measurements in the same $\sqrt{s_{NN}}$ range. Any statistically significant departure of the measured $\alpha(\sqrt{s_{NN}})$ or $R_\mathrm{diff}^2(\sqrt{s_{NN}})$ from the smooth, monotonic hadronic baselines established in this work would provide compelling evidence for dynamics beyond pure hadronic transport, potentially signalling the proximity of the QCD critical point.

\backmatter


\bmhead{Acknowledgements}
The authors would like to thank B. Pórfy, and M. Nagy for the insightful discussions. We acknowledge the support from the NKFIH grants PD-146589 and NKKP ADVANCED 152097.



\bmhead{Data Availability Statement}
This manuscript has data included as supplementary material. [Author's comment: The datasets generated during the current study are available at \href{https://doi.org/10.5281/zenodo.23036245}{\nolinkurl{https://doi.org/10.5281/zenodo.23036245}}].
\bmhead{Code Availability Statement}
This manuscript has associated code/software in a data repository. [Author's comment: The code used for developing the analysis code of the current study (\href{https://github.com/molnarmatyas/UrQMD_EPJApaper_analysisarchive}{\nolinkurl{https://github.com/molnarmatyas/UrQMD\_EPJApaper\_analysisarchive}}) is available in the GitHub repository, \href{https://github.com/csanadm/LevySourceFit}{\nolinkurl{https://github.com/csanadm/LevySourceFit}}].
\bmhead{Open Access}
This article is licensed under a Creative Commons Attribution 4.0 International License, which permits use, sharing, adaptation, distribution and reproduction in any medium or format, as long as you give appropriate credit to the original author(s) and the source, provide a link to the Creative Commons licence, and indicate if changes were made. The images or other third party material in this article are included in the article's Creative Commons licence, unless indicated otherwise in a credit line to the material. If material not included in the article's Creative Commons licence and your intended use is not permitted by statutory regulation or exceeds the permitted use, you will need to obtain permission directly from the copyright holder. To view a copy of this licence, visit \href{http://creativecommons.org/licenses/by/4.0/.}{http://creativecommons.org/licenses/by/4.0/}.


\begin{appendices}
\section{~}\label{secA1}
In this appendix, Table~\ref{tab:sysunc_full} lists the average relative systematic uncertainties.
\begin{table*}[!htpb]
\centering
\fontsize{7.5pt}{8.2pt}\selectfont
\caption{Average relative systematic uncertainties of the emission source parameters from different uncertainty sources at all collision energies, in percent. \label{tab:sysunc_full}}
\setlength{\tabcolsep}{10pt}
\renewcommand{\arraystretch}{1.1}
\begin{tabular}{llccccc}
\toprule
$\sqrt{s_{NN}}$ & \textbf{Source} & $\alpha$ & $\lambda^{*}$ & $R_{\mathrm{out}}$ & $R_\mathrm{side}$ & $R_\mathrm{long}$\\
\midrule
\multirow[m]{4}{*}{3.0 GeV} & $Q_\mathrm{LCMS}^\mathrm{max}$ &  +0.35 / -0.65 &  +0.25 / -0.14 &  +1.07 / -1.22 &  +2.26 / -2.38 &  +3.39 / -3.90\\
 &  $\rho_\mathrm{fit}^\mathrm{max}$ &  +0.55 / -1.83 &  +0.93 / -0.29 &  +0.51 / -0.10 &  +0.23 / -0.30 &  +0.53 / -1.54\\
 & $N_{\rm events}$ &  +0.02 / -0.14 &  +0.04 / -0.01 &  +0.00 / -0.04 &  +0.01 / -0.02 &  +0.02 / -0.12\\
 & Total &  +0.75 / -1.98 &  +0.99 / -0.34 &  +1.36 / -1.26 &  +2.27 / -2.41 &  +3.43 / -4.37\\
\midrule
\multirow[m]{4}{*}{3.2 GeV} & $Q_\mathrm{LCMS}^\mathrm{max}$ &  +0.31 / -0.52 &  +0.19 / -0.06 &  +1.07 / -1.06 &  +2.21 / -2.71 &  +3.62 / -4.00\\
 & $\rho_\mathrm{fit}^\mathrm{max}$ &  +0.52 / -1.57 &  +0.81 / -0.27 &  +0.48 / -0.11 &  +0.17 / -0.31 &  +0.52 / -1.32\\
 & $N_{\rm events}$ &  +0.05 / -0.14 &  +0.05 / -0.01 &  +0.04 / -0.06 &  +0.01 / -0.02 &  +0.06 / -0.11\\
 & Total &  +0.70 / -1.68 &  +0.85 / -0.29 &  +1.34 / -1.12 &  +2.22 / -2.74 &  +3.67 / -4.24\\
\midrule
\multirow[m]{4}{*}{3.5 GeV} & $Q_\mathrm{LCMS}^\mathrm{max}$ &  +0.32 / -0.58 &  +0.10 / -0.07 &  +0.96 / -1.10 &  +2.26 / -2.50 &  +3.58 / -4.45\\
 & $\rho_\mathrm{fit}^\mathrm{max}$ &  +0.34 / -1.34 &  +0.69 / -0.18 &  +0.48 / -0.10 &  +0.11 / -0.27 &  +0.34 / -1.16\\
 & $N_{\rm events}$ &  +0.04 / -1.13 &  +0.54 / -0.02 &  +0.01 / -0.21 &  +0.02 / -0.04 &  +0.04 / -0.71\\
 & Total &  +0.57 / -2.00 &  +0.94 / -0.20 &  +1.24 / -1.22 &  +2.26 / -2.52 &  +3.60 / -4.71\\
\midrule
\multirow[m]{4}{*}{3.9 GeV} & $Q_\mathrm{LCMS}^\mathrm{max}$ &  +0.37 / -0.40 &  +0.19 / -0.05 &  +0.88 / -0.93 &  +2.16 / -2.60 &  +3.67 / -4.22\\
 & $\rho_\mathrm{fit}^\mathrm{max}$ &  +0.23 / -0.96 &  +0.52 / -0.13 &  +0.45 / -0.08 &  +0.08 / -0.19 &  +0.25 / -0.93\\
 & $N_{\rm events}$ &  +0.01 / -0.46 &  +0.19 / -0.00 &  +0.00 / -0.13 &  +0.00 / -0.04 &  +0.00 / -0.29\\
 & Total &  +0.54 / -1.25 &  +0.63 / -0.16 &  +1.12 / -0.98 &  +2.17 / -2.60 &  +3.69 / -4.39\\
\midrule
\multirow[m]{4}{*}{4.2 GeV} & $Q_\mathrm{LCMS}^\mathrm{max}$ &  +0.26 / -1.45 &  +1.00 / -0.21 &  +0.87 / -1.05 &  +2.32 / -2.56 &  +3.95 / -5.29\\
 & $\rho_\mathrm{fit}^\mathrm{max}$ &  +0.03 / -0.43 &  +0.24 / -0.02 &  +0.49 / -0.12 &  +0.02 / -0.13 &  +0.10 / -0.56\\
 & $N_{\rm events}$ &  +0.31 / -0.71 &  +0.36 / -0.17 &  +0.10 / -0.09 &  +0.03 / -0.04 &  +0.24 / -0.41\\
 & Total &  +0.51 / -1.73 &  +1.12 / -0.28 &  +1.13 / -1.08 &  +2.32 / -2.56 &  +3.97 / -5.35\\
\midrule
\multirow[m]{4}{*}{4.5 GeV} & $Q_\mathrm{LCMS}^\mathrm{max}$ &  +0.24 / -1.10 &  +0.78 / -0.15 &  +0.81 / -0.69 &  +2.21 / -2.54 &  +3.88 / -5.10\\
 & $\rho_\mathrm{fit}^\mathrm{max}$ &  +0.05 / -0.41 &  +0.24 / -0.03 &  +0.42 / -0.11 &  +0.02 / -0.14 &  +0.06 / -0.53\\
 & $N_{\rm events}$ &  +0.17 / -0.48 &  +0.26 / -0.09 &  +0.05 / -0.08 &  +0.02 / -0.03 &  +0.12 / -0.31\\
 & Total &  +0.41 / -1.32 &  +0.88 / -0.20 &  +1.03 / -0.75 &  +2.21 / -2.55 &  +3.89 / -5.15\\
\midrule
\multirow[m]{4}{*}{5.2 GeV} & $Q_\mathrm{LCMS}^\mathrm{max}$ &  +0.22 / -1.98 &  +1.36 / -0.19 &  +0.73 / -0.74 &  +2.21 / -2.40 &  +4.14 / -5.61\\
 & $\rho_\mathrm{fit}^\mathrm{max}$ &  +0.08 / -0.28 &  +0.14 / -0.03 &  +0.38 / -0.13 &  +0.02 / -0.11 &  +0.02 / -0.32\\
 & $N_{\rm events}$ &  +0.23 / -1.10 &  +0.58 / -0.13 &  +0.05 / -0.12 &  +0.02 / -0.06 &  +0.17 / -0.62\\
 & Total &  +0.43 / -2.35 &  +1.51 / -0.25 &  +0.92 / -0.82 &  +2.21 / -2.41 &  +4.15 / -5.69\\
\midrule
\multirow[m]{4}{*}{6.2 GeV} & $Q_\mathrm{LCMS}^\mathrm{max}$ &  +0.16 / -1.16 &  +0.92 / -0.11 &  +0.66 / -0.71 &  +2.19 / -2.47 &  +4.14 / -5.28\\
 & $\rho_\mathrm{fit}^\mathrm{max}$ &  +0.17 / -0.30 &  +0.14 / -0.06 &  +0.28 / -0.12 &  +0.04 / -0.09 &  +0.05 / -0.25\\
 & $N_{\rm events}$ &  +0.08 / -0.58 &  +0.31 / -0.05 &  +0.03 / -0.09 &  +0.01 / -0.06 &  +0.06 / -0.36\\
 & Total &  +0.31 / -1.41 &  +1.01 / -0.17 &  +0.78 / -0.75 &  +2.19 / -2.48 &  +4.15 / -5.31\\
\midrule
\multirow[m]{4}{*}{7.2 GeV} & $Q_\mathrm{LCMS}^\mathrm{max}$ &  +0.19 / -0.73 &  +0.72 / -0.08 &  +0.64 / -0.72 &  +2.16 / -2.53 &  +4.15 / -5.25\\
 & $\rho_\mathrm{fit}^\mathrm{max}$ &  +0.21 / -0.26 &  +0.11 / -0.08 &  +0.23 / -0.11 &  +0.05 / -0.08 &  +0.06 / -0.21\\
 & $N_{\rm events}$ &  +0.03 / -0.33 &  +0.18 / -0.02 &  +0.01 / -0.06 &  +0.01 / -0.05 &  +0.02 / -0.21\\
 & Total &  +0.32 / -0.91 &  +0.77 / -0.14 &  +0.72 / -0.74 &  +2.16 / -2.53 &  +4.15 / -5.27\\
\midrule
\multirow[m]{4}{*}{7.7 GeV} & $Q_\mathrm{LCMS}^\mathrm{max}$ &  +0.18 / -0.69 &  +0.59 / -0.07 &  +0.69 / -0.74 &  +2.19 / -2.49 &  +4.21 / -5.05\\
 & $\rho_\mathrm{fit}^\mathrm{max}$ &  +0.20 / -0.24 &  +0.11 / -0.08 &  +0.23 / -0.10 &  +0.04 / -0.10 &  +0.05 / -0.23\\
 & $N_{\rm events}$ &  +0.02 / -0.25 &  +0.14 / -0.01 &  +0.01 / -0.06 &  +0.01 / -0.04 &  +0.01 / -0.17\\
 & Total &  +0.29 / -0.85 &  +0.65 / -0.13 &  +0.76 / -0.77 &  +2.19 / -2.50 &  +4.21 / -5.07\\
\midrule
\multirow[m]{4}{*}{9.2 GeV} & $Q_\mathrm{LCMS}^\mathrm{max}$ &  +0.16 / -1.71 &  +1.27 / -0.12 &  +0.59 / -0.52 &  +2.20 / -2.44 &  +4.21 / -5.54\\
 & $\rho_\mathrm{fit}^\mathrm{max}$ &  +0.31 / -0.22 &  +0.09 / -0.11 &  +0.20 / -0.10 &  +0.05 / -0.03 &  +0.04 / -0.07\\
 & $N_{\rm events}$ &  +0.13 / -0.48 &  +0.25 / -0.06 &  +0.03 / -0.06 &  +0.04 / -0.07 &  +0.09 / -0.27\\
 & Total &  +0.45 / -1.85 &  +1.32 / -0.20 &  +0.66 / -0.58 &  +2.20 / -2.44 &  +4.22 / -5.56\\
\midrule
\multirow[m]{4}{*}{11.5 GeV} & $Q_\mathrm{LCMS}^\mathrm{max}$ &  +0.17 / -0.94 &  +0.79 / -0.06 &  +0.56 / -0.68 &  +2.19 / -2.54 &  +4.23 / -5.38\\
 & $\rho_\mathrm{fit}^\mathrm{max}$ &  +0.22 / -0.17 &  +0.06 / -0.08 &  +0.14 / -0.09 &  +0.04 / -0.03 &  +0.05 / -0.09\\
 & $N_{\rm events}$ &  +0.06 / -0.30 &  +0.15 / -0.03 &  +0.01 / -0.05 &  +0.02 / -0.05 &  +0.05 / -0.17\\
 & Total &  +0.33 / -1.07 &  +0.83 / -0.13 &  +0.60 / -0.70 &  +2.19 / -2.54 &  +4.23 / -5.39\\
\midrule
\multirow[m]{4}{*}{14.5 GeV} & $Q_\mathrm{LCMS}^\mathrm{max}$ &  +0.18 / -0.76 &  +0.45 / -0.02 &  +0.56 / -0.62 &  +2.14 / -2.59 &  +4.16 / -5.11\\
 & $\rho_\mathrm{fit}^\mathrm{max}$ &  +0.17 / -0.13 &  +0.05 / -0.06 &  +0.12 / -0.08 &  +0.03 / -0.03 &  +0.06 / -0.09\\
 & $N_{\rm events}$ &  +0.03 / -0.19 &  +0.10 / -0.02 &  +0.02 / -0.04 &  +0.02 / -0.04 &  +0.02 / -0.12\\
 & Total &  +0.30 / -0.85 &  +0.48 / -0.08 &  +0.59 / -0.64 &  +2.14 / -2.59 &  +4.16 / -5.11\\
\midrule
\multirow[m]{4}{*}{19.6 GeV} & $Q_\mathrm{LCMS}^\mathrm{max}$ &  +0.16 / -1.30 &  +0.76 / -0.07 &  +0.60 / -0.60 &  +2.21 / -2.64 &  +4.30 / -5.35\\
 & $\rho_\mathrm{fit}^\mathrm{max}$ &  +0.17 / -0.18 &  +0.06 / -0.06 &  +0.13 / -0.08 &  +0.03 / -0.04 &  +0.06 / -0.06\\
 & $N_{\rm events}$ &  +0.10 / -0.40 &  +0.18 / -0.05 &  +0.03 / -0.05 &  +0.03 / -0.10 &  +0.07 / -0.20\\
 & Total &  +0.32 / -1.45 &  +0.81 / -0.12 &  +0.63 / -0.62 &  +2.21 / -2.64 &  +4.30 / -5.36\\
\midrule
\multirow[m]{4}{*}{27.0 GeV} & $Q_\mathrm{LCMS}^\mathrm{max}$ &  +0.14 / -1.11 &  +0.49 / -0.05 &  +0.53 / -0.55 &  +2.20 / -2.66 &  +4.19 / -5.14\\
 & $\rho_\mathrm{fit}^\mathrm{max}$ &  +0.11 / -0.20 &  +0.06 / -0.03 &  +0.08 / -0.07 &  +0.02 / -0.07 &  +0.04 / -0.08\\
 & $N_{\rm events}$ &  +0.10 / -0.32 &  +0.13 / -0.04 &  +0.02 / -0.04 &  +0.04 / -0.11 &  +0.05 / -0.18\\
 & Total &  +0.30 / -1.25 &  +0.53 / -0.09 &  +0.55 / -0.56 &  +2.20 / -2.67 &  +4.20 / -5.14\\
\bottomrule
\end{tabular}
\end{table*}




\end{appendices}

\bibliography{references.bib}

\end{document}